\documentclass[12pt]{article} 
\begin{document} 
%

\title{{\bf{Time-variation of the fine structure constant
and mass-energy conservation.}}
\author{G.R. Filewood\\
Research Centre for High Energy Physics,\\
School of Physics,
University of Melbourne,\\
Parkville, Victoria 3052 Australia.}
\date{\today}
\abstract
A brief note on the issue of possible time-variation of 
the fine structure constant $\alpha_{em}$;
it is shown that
such a variation should  violate mass-energy conservation
in the absence of proton decay.\\ }

\maketitle

There has been some recent interest (see for example \cite{Webb},
\cite{Oberhummer} and \cite{Ivanchik} and for an excellent 
review of the subject area \cite{Uzan}) in the possibility 
of time-variation of the fine structure constant;
the coupling of the electromagnetic force.

This brief note is to raise a simple point which seems not 
to have been noted  by previous authors and which has 
implications for the first law of thermodynamics; one
of the most cherished (and most strongly
empirically validated) principles in physics.

Currently there is no experimental evidence of proton
decay. Whilst this does not prove that the proton is
absolutely stable, in the absence of any empirical evidence
to the contrary in spite of dedicated searches it is the
only assumption that can claim any empirical validity. The same 
comment applies to the electron.

By the time/Energy 
 uncertainty relation ${\Delta}t.{\Delta}E
\geq{\hbar\over2}$ we can, 
theoretically in principle, measure the
rest mass of an {\it{absolutely}} stable, and only an
absolutely stable,  particle to any desired
degree of accuracy. (Particles that 
have a finite lifetime must 
always have an uncertainty in their mass). Now, in  the 
absence of a complete knowledge of the functional
  dependence of the masses of
the electron and the proton on dynamical
 factors we cannot 
pinpoint the exact way in which these masses may vary
 with a time-varying
$\alpha_{em}$ but, because both carry 
electromagnetic charge and have
electromagnetic self-interaction, we 
know that the masses in question must
have a functional dependence on the value
 of $\alpha_{em}$ and should
therefore vary if $\alpha_{em}$ varies. Of course one may
hypothesise that, with a time-varying $\alpha_{em}$, both masses 
vary in the same way and that the ratio of the
masses does not change. That this cannot be the
 case can be understood from the 
following. We can express some of the factors 
(but not necessarily all) which must influence
the rest masses of the proton and electron 
in terms of the following list of the forces in
nature which have an input into the respective
mass generation;

\[
M_p;\{{\alpha}_{s},\;{\alpha}_{em},\;{\alpha}_{w},\;G_N.\}
\;\;\mbox{and}\;\;
M_e;\{{\alpha}_{em},\;
{\alpha}_{w},\;G_N.\}
\]

\noindent
i.e. we don't need a complete knowledge of the origin of the
masses in question to assert which of the
 known forces must make
a contribution to the mass. Here $\alpha_s$ is of course the
coupling of the strong interaction and the other couplings
are the weak and gravitational interaction.
We know empirically that the
electron, like other leptons, does not
 experience the strong force so
the $\alpha_s$ cannot contribute to the electron rest mass.

Therefore in the event of a continuous 
time-varying ${\alpha}_{em}$ there must be a
continuous fine-tuning of ${\alpha}_s$ or the ratio of 
rest masses ${M_p}/{M_e}$ will change. Such a 
continuous fine-tuning is quite untenable since
the structural components upon which
and through which  the two couplings
act are quite different; the gluon interaction is 
quite a different process to 
electromagnetism and the dynamics
of the quark-gluon interaction 
is quite distinct from
the electromagnetic interaction
so that the  mass components 
generated by the two different
forces will not have any simple 
functional inter-relationship.
Moreover,
mechanisms which we
might reasonably suppose could 
account for a time-varying
${\alpha}_{em}$ seem inappropriate
 in the case of the
strong interaction; we might, for 
example (as I think Dirac
proposed - or was it in relation the
the gravitational constant as well?)
 consider that the speed of light is some sort of
function of the diameter / age of the 
universe and suppose that
${\alpha}_{em}$ varies accordingly? 
Since electromagnetism is 
a long-range force perhaps this is a 
reasonable speculation
but it is difficult to see how a 
parameter such as ${\alpha}_s$,
which is a short-range force 
confined in its' own `colour-space', 
could conform to such a scheme or 
any similar scenario. (That 
the running
couplings follow different paths in 
phase space below the
unification scale  need not 
necessarily  be an issue here since
the measurement in question is 
that of the rest mass of the proton and
electron which, by definition, 
is the low-energy limit; put
another way,  the {\it{time}} 
variation of the couplings 
at a fixed energy scale and the
{\it{energy scale}} variation of
the couplings
are quite separate issues physically). If 
any fixed energy scale 
time variation of ${\alpha}_s$ and ${\alpha}_{em}$
results from physically independent mechanisms
it is inconceivable that the requisite fine-tuning
required to maintain a fixed $M_p/M_e$ should occur.

But then what is wrong with a 
time-varying ${M_p}/{M_e}$ rest
mass ratio? The answer is that, 
if both these particles are 
absolutely stable, any variation of the ratio is a
violation of mass-energy conservation. 
Mass simply cannot
`disappear' and the first law of 
thermodynamics remain intact.

It is suggested therefore that, should good 
empirical evidence  be obtained of a 
time-varying ${\alpha}_{em}$, 
physics must consider dumping the
first law of thermodynamics. 
Alternatively, and possibly more
fruitfully, alternative explanations for
 the empirical observations
should be sought. Equivalently it 
is suggested that, if 
mass-energy conservation is indeed 
preserved in the universe, and this
certainly seems to be the case 
empirically, and if the proton and
electron are indeed stable then 
${\alpha}_{em;q^2=0}$ is an invariant
of nature. By the same argument 
${\alpha}_s$ at some fixed energy scale
and $\alpha_w$ must obey the same 
principle. The gravitational coupling
constant however is not so constrained 
since it couples to all energy and
matter equally thus  it is possible 
$G_N$  may vary
in time  without variation of
the ${M_p}/{M_e}$ rest mass ratio.


\begin{thebibliography}{9}

\bibitem{Webb}
J Webb et. al. astro-ph/0210299 and 
astro-ph/0210531

\bibitem{Oberhummer}
H Oberhummer et. al. astro-ph/0210459

\bibitem{Ivanchik}
A Ivanchik et.al. astro-ph/0210299

\bibitem{Uzan}J Uzan.
hep-ph/0205340  and contained references.

\end{thebibliography}
\end{document}